\documentclass[12pt,preprint]{aastex}
\usepackage{epsfig}

\begin{document}

\shortauthors{Vanderlinde, Levine, \& Rappaport}
\shorttitle{Orbital Period of Scorpius X-1}

\title{ {\it RXTE} All-Sky Monitor Detection of the Orbital Period of
 Scorpius X-1}

\author{Keith W. Vanderlinde, Alan M. Levine, and Saul A. Rappaport}

\affil{Department of Physics and Center for Space Research}
\affil{Massachusetts Institute of Technology}
\affil{Room 37-575, 77 Massachusetts Ave., Cambridge, MA 02139}
\email{aml@space.mit.edu}

\begin{abstract}

The orbital period of Scorpius X-1 has been accepted as 0.787313 d
since its discovery in archival optical photometric data by Gottlieb,
Wright \& Liller (1975). This period has apparently been confirmed
multiple times in the years since in both photometric and
spectroscopic optical observations, though to date only marginal
evidence has been reported for modulation of the X-ray intensity at
that period. We have used data taken with the All Sky Monitor on board
the {\it Rossi X-Ray Timing Explorer } over the past 6 years to search
for such a modulation.  A major difficulty in detecting the orbit in
X-ray data is presented by the flaring behavior in this source,
wherein the (1.5-12 keV) X-ray intensity can change by up to a factor
of two within a fraction of a day.  These flares contribute nearly
white noise to Fourier transforms of the intensity time series, and
thereby tend to obscure weak modulations, i.e., of a few percent or
less.  We present herein a new technique for substantially reducing
the effects of the flaring behavior while, at the same time, retaining
much of any periodic orbital modulation, provided only that the two
temporal behaviors exhibit different spectral signatures. Through such
a search, we have found evidence for orbital modulation at the $\sim
1\%$ level with a period of 0.78893 d, equal within our accuracy to a
period which differs by 1 cycle per year from the accepted value. If
we compare our results with the period of the 1 year sideband cited by
Gottlieb et al. we conclude that the actual period may in fact be
0.78901 d.  Finally, we note that many of the reported optical
observations of Sco X-1 have been made within one or two months of
early June, when Sco X-1 transits the meridian at midnight.  All
periodicity searches based only on such observations would have been
subject to the same 1 cycle per year alias that affected the search of
\citet{GWL75}.

\end{abstract}

\keywords{binaries: close---stars: individual (Sco X-1)---X-rays:
binaries }

\section{Introduction}

Scorpius X-1 is the most prominent of the bright Galactic bulge X-ray
sources. These are persistent, high-luminosity X-ray sources located
in directions more or less towards the Galactic center, and are
typified by, in addition to Sco X-1, GX340+0, GX349+2, GX3+1, GX5--1,
GX9+1, GX9+9, GX13+1, and GX17+2.  The 2 to 10 keV intensities of
these sources are often seen to vary within a range limited to about a
factor of two or so; much larger changes in strength are rare.
Neither eclipses nor persistent periodic pulsations have been detected
in any of them.  These sources are believed to be low-mass X-ray
binaries (LMXBs), binaries with short orbital periods in which
accretion onto a neutron star primary is fed by Roche lobe overflow
from a companion less massive than the Sun.  Two classes of LMXBs, the
``Z'' and ``atoll'' types, are distinguished by distinctive tracks
seen in X-ray color-color diagrams.  In both types of source, the
overall intensity and features of Fourier power density spectra,
including quasiperiodic oscillations, also change in a characteristic
way with location along the color-color track \citep{Has89,vdK00}.
The brightest bulge sources are of the Z-type, which may be emitting
close to the Eddington limit for a neutron star of mass $\sim
1.4~M_\odot$. Although Sco X-1 is by far the brightest persistent
source, it is not likely to be intrinsically more luminous than the
other Z sources; rather it is thought to be $\sim 3$ times closer than
most of the bulge sources \citep{BFG99}.

Sco X-1 exhibits only the so-called normal and flaring branches of the
Z-source X-ray color-color diagram; it does not exhibit a ``horizontal
branch'' \citep{Has89}.  During frequent excursions onto the flaring
branch, the source intensity increases by up to a factor of $\sim 2$.
These flares often come and go on time scales of hours \citep[and
references therein]{Bra75}.

The high level of activity that Sco X-1 displays in X-rays also
appears in the optical and radio bands where the variability is
correlated with the variability in the X-ray band
\citep{San66,Bra75,Can75,WGL75}.  In the optical, it ranges
over blue magnitudes of 11.8 to 13.6, but on time scales of 1 day or
less it rarely changes by more than 1 magnitude.

Apart from strong X-ray emission and occasional weak radio emission,
the binary nature of the bright Galactic bulge sources is only subtly,
if at all, manifested observationally.  Of the 9 sources listed above,
orbital periods have been determined for only Sco X-1, GX 9+9, and,
possibly, GX349+2 (Gottlieb, Wright \& Liller 1975, hereafter GWL;
Hertz \& Wood 1988; Schaefer 1990; Wachter \& Margon 1996).

The orbital period of Sco X-1 was first found by GWL in a search for
periodicities in 85 years of photometric optical data obtained from
plates in the Harvard collection. GWL found four distinct candidates
for the orbital period, viz., 0.787313 d, 0.78901 d, 0.81069 d, and
3.74001 d. They identified 0.787313 d as the likely orbital period,
while the other periods were explained as one year, one month, and one
day sidebands, respectively, i.e., as artifacts of the limited
coverage of the observations.  Use of a large number of measurements
was critical to GWL's success in elucidating these periodicities
against the background of high frequency noise from random variations
in brightness; in the optical the orbital modulation has a full
amplitude of only $\sim 0.2$ magnitude \citep{GWL75,WGL75}.

\citet{CC75} carried out spectroscopic observations of Sco X-1 and
found radial velocity variations in the He II and H emission lines
with a period of $0.787 \pm 0.006$ d and thereby conclusively
demonstrated the orbital nature of the 0.787 day variations.  If the
emission lines are formed, e.g., in an accretion disk around the X-ray
emitting star, then the epoch of minimum brightness found by GWL
corresponds to superior conjunction of the X-ray source
\citep{CC75,Cra76}.

The presence of a 0.787 d period has been confirmed in a number of
other studies based on optical photometry (e.g., van Genderen 1977;
see Augusteijn et al. 1992) or on radial velocities obtained from
optical spectroscopy (e.g., Bord et al. 1976; LaSala \& Thorstensen
1985).  Further information on the binary system has recently been
obtained by \citet{SC02}, who used optical spectroscopy to uncover
evidence of the donor star in the binary system and to obtain
estimates of its radial velocity amplitude.

No significant periodicity at 0.787313 d or any nearby period in the
X-ray flux from Sco X-1 has been reported.  \citet{PH87}
searched for a signal at the GWL period in 5 years of 3-6 keV
intensity measurements from the All-Sky Monitor on {\it Ariel V } and
found only a weak modulation with an amplitude of $\sim 0.4$\%.
Priedhorsky \& Holt did not report any investigations of variability
at other periods.

We have analyzed X-ray intensities of Sco X-1 recorded over more than
six years by the All Sky Monitor (ASM) on board the {\it Rossi X-ray
Timing Explorer}, and have found evidence of a periodicity with a
period of 0.78893 d. This period is consistent with GWL's 0.78901 d
period, and we therefore tentatively propose that the latter value is
the orbital period of Sco X-1, while the 0.787313 d period is a
one-year sideband produced as a consequence of the observational
window function.

\section{Observations}

The ASM comprises 3 coded-aperture cameras, or Scanning Shadow Cameras
(SSCs), mounted as an assembly that can be rotated around a single
axis by a motorized drive \citep{lev96}. The shadow pattern data from
each 90-second exposure with a single SSC are analyzed to obtain
intensity estimates of each pre-cataloged source that is within the
field of view.  The analysis is done for the overall energy band of
the ASM, $\sim$ 1.5-12 keV, as well as the nominal 1.5-3 keV, 3-5 keV,
and 5-12 keV bands (the A, B, and C bands, respectively).

Our analysis involves both intensities and colors, i.e., ratios of
intensities from different energy bands.  We utilized the intensities
produced as part of the standard processing done by the {\it RXTE }
ASM team \citep{lev96}.  In the derivation of these intensity
estimates, corrections are made for instrumental characteristics which
have changed over time.  The correction factors are constructed to
remove trends in the derived intensities of the Crab Nebula (including
the pulsar) over the course of the mission and to remove any
dependence on SSC, position in the field of view, etc.  The factors
obtained for the Crab Nebula are also applied to other sources even
though this is an approximation that becomes less accurate as the
source spectrum deviates more strongly from that of the Crab.  One may
therefore expect some systematic errors in the intensities of Sco X-1
since its spectrum is not the same as that of the Crab; these are
expected to be small in the intensities derived from SSC 2 and larger
in the intensities derived from SSCs 1 and 3.  The gas-filled
proportional counter detector in SSC 1 has a slow leak that has
resulted in the photon energy to pulse height conversion gain
increasing at a rate of approximately 9\% per year.  On one occasion,
early in the year 2000, the flight software pulse height intervals
that define the individual energy bands of SSC 1 were modified to
reduce the effects of the gain changes.  For the present analysis, we
have made additional empirical corrections to the A and B channel
intensities from observations with SSC 1 to remove much of the
remaining signature of instrumental gain changes.  In SSC 3, 2 of 8
carbon-coated quartz fiber anodes were irretrievably damaged in the
first week after launch, and a third anode failed after about 7
months.  Gross calibration changes since then have rendered 2 of the
remaining 5 anodes unusable.  Because of these complications, we do
not use data from SSC 3 in the present analysis.  In all cases,
however, the gain changes and correction factors should not contribute
any significant amount of spurious power at temporal frequencies close
to 1 cycle d$^{-1}$ where the search for orbital modulations takes
place.

Over 11,000 intensity measurements of Sco X-1 have been obtained by
each of SSCs 1 and 2 since the ASM began routine operations in late
February 1996.  The 1.5-12 keV light curve of Sco X-1 using data from
SSC 2 only is presented in Figure 1a. The flaring behavior associated
with excursions onto the flaring branch is clearly visible, with
frequent and irregular flares above a relatively stable baseline. In
Figure 1b we show the hardness ratio I(5-12 keV)/I(3-5 keV). Note, in
particular, that the hardness ratio increases strongly during the
flares \citep{Has89,vdK00}.

\section{Analysis}

In this section we first describe a standard Fourier transform
analysis of the time series of X-ray intensity data obtained from Sco
X-1. The prominent flaring activity is shown to introduce a
significant amount of noise into the transform which will likely mask
any signature of a weak orbital modulation.  We then introduce a new
analysis algorithm which greatly reduces the effects of the flaring
behavior.  This approach makes use of the strong spectral changes
during the flares.

To begin our analysis, the intensities from each of SSCs 1 and 2 were
assigned to 0.01-d time bins, and the mean intensity was subtracted
from the populated bins. Each resulting linear array was extended by a
factor of eight by padding with zeros and was then transformed using a
Fast Fourier Transform (hereafter FFT) code.  The power at each
frequency (the quadratic sum of the real and imaginary parts of the
output of the FFT) was divided by the mean power of the entire
spectrum.  The extension of the linear arrays yields interpolated
power density spectra (PDSs) in which the sensitivity for detection of
periodicities is more uniform than in an uninterpolated PDS.  The
results from SSC 2 are shown in Figure 2a, and provide no obvious
evidence for any significant periodicity.  Even modulations with
frequencies close to one cycle per day, which often appear in PDSs of
ASM data as a result of the observational window function acting on
low-frequency source variability, are not noticeable in the PDS of the
raw data. This is taken as evidence that the power density is
dominated by the effects of the flares in the intensity of Sco X-1,
and thus any small amplitude modulations due to the orbit are likely
to be undetectable as well.

In order to significantly reduce the noise power introduced by the
flaring behavior and possibly increase the sensitivity for detection
of weak modulation at the orbital frequency, we utilize the fact that
the intensity of Sco X-1, like other Z sources, exhibits distinct
spectral hardening during the flares (see, e.g., Hasinger \& van der
Klis, 1989). The relation between intensity and spectral hardness from
the ASM data set itself is illustrated in Figure 3, where the
intensity in each of the A, B, and C bands, which we will denote by
$I_A$, $I_B$, and $I_C$, respectively, is plotted vs. the hardness
ratio defined as $R_{\rm C,B} = I_C/I_B$. Note that on the average the
intensity increases monotonically, though not quite linearly, as
$R_{\rm C,B}$ increases.  The smooth curves in each panel illustrate
models of the spectral behavior of the flares that we utilize below.

Our flare-suppression algorithm utilizes simple analytic functions to
describe the mean intensity--hardness-ratio behavior.  These
expressions are then used to ``correct'' the intensity history of Sco
X-1 so as to largely remove the flaring behavior.  In the following
detailed description of the method, we use unprimed symbols to denote
quantities that would be observed in the hypothetical {\em absence} of
any modulations due to the orbit, and primed symbols to denote
quantities affected by orbital modulations.

We start by defining the {\em mean} spectral track that would be
followed by the source in an intensity--hardness diagram, such as
those shown in Figure 3, in the absence of orbital modulations by:
\begin{equation}
\hat{I}_k(t) = F_k\left[R_{ij}(t)\right]    ,
\end{equation}
where $\hat{I}_k(t)$ is the model intensity in the $k$th energy
channel, $t$ is the elapsed time, and $F_k$ describes the shape of the
track which is taken to be a function of the hardness ratio in two
energy channels $i$ and $j$:
\begin{equation}
R_{ij}(t) \equiv \frac{I_i(t)}{I_j(t)}
\end{equation}
with channel $i$ being at higher energies than channel $j$. Here, the
energy channel $k$ whose intensity we are modeling may or may not be
the same as channel $i$ or $j$.  The intensities $I_k(t)$, model
intensities $\hat{I}_k(t)$, and hardness ratios $R_{ij}(t)$ are
defined as those that would be detected if there were no modulations
due to the orbit.

When any orbital modulation is defined so as to have a mean value of
zero and is small in amplitude, we can still describe the mean
intensity--hardness relation using the same function $F_k$:
\begin{equation}
\hat{I}_k'(t) \simeq F_k\left[R_{ij}'(t)\right]   ,
\end{equation}
where, as stated above, the primes indicate hardness ratios and model
intensities affected by orbital modulations.  We model the orbital
modulation in the $k$th channel by a simple multiplicative term,
$M_k(t)$, to represent a small increase or decrease from the mean in
the degree of scattering or absorption out of the line of sight, or
scattering into the line of sight. Thus,
\begin{equation}
R_{ij}'(t) \simeq \frac{I_i(t) M_i(t)}{I_j(t) M_j(t)} \, .
\end{equation}
A modified data set, $C_k(t)$, approximately corrected for
the flaring, is then given by:
\begin{equation}
C_k(t) \equiv \frac{I'_k(t)}{F_k\left[R_{ij}'(t)\right]}
\simeq \frac{I_k(t)M_k(t)}{F_k\left[R_{ij}(t)M_i(t)/M_j(t)\right]} \, ,
\end{equation}
where the first ratio is expressed in terms of measured quantities
(i.e., with the orbital modulation included), and the second ratio is
expressed in terms of the intrinsic source behavior.

In order to learn about the properties of the corrected data set, we
expand the numerator and denominator in Taylor series about the point
$M_i = M_j = 1$, i.e., we take $M(t) = 1 + \delta M(t)$.  The above
expression can then be written:
\begin{equation}
C_k \simeq \frac{I_k (1+\delta M_k)}{\left[F_k(R_{ij})+
\left(\frac{\partial F_k}{\partial R_{ij}'}\right)_0\left(
\frac{\partial R_{ij}'}{\partial M_i}\right)_0\delta M_i
+\left(\frac{\partial F_k}{\partial R_{ij}'}\right)_0\left(
\frac{\partial R_{ij}'}{\partial M_j}\right)_0\delta M_j
\right]} \,  ,
\end{equation}
where we have dropped explicit time dependences in
the interest of simplicity.  After collecting terms and making use
of the assumption that the $\delta M'$s are small, we find the
following simplified expression for the behavior of the corrected
data set in the presence of orbital modulations:
\begin{equation}
C_k \simeq \frac{I_k}{F_k(R_{ij})}\left\{1+\delta M_k -
\frac{d\ln F_k}{d\ln R_{ij}}\left[\delta M_i - \delta M_j\right]
\right\}  \,  .
\end{equation}
To the extent that the flaring behavior follows a distinct track in the
color--intensity diagram, the factor $I_k/F_k(R_{\rm i,j})$ is nearly
constant, i.e., independent of flaring behavior.  This factor will also
contain some residual random fluctuations in the source intensity that do
not follow the color--intensity track, as well as any measurement errors.
The factor in curly brackets contains the desired time-dependent orbital
modulations, i.e., the $\delta M'$s.

From equation (7) one may see that the manifestation of orbital
modulations in the corrected data set consists of two parts: (i)
the modulation in the channel being corrected, and (ii) a
term proportional to the difference in the orbital modulation in
the two channels used to define the hardness ratio.  There are
a couple of noteworthy combinations of terms in this equation.
First, if $\delta M_i = \delta M_j$, the modulation appearing
in $C_k$ is just that of the orbital modulation in the $k$th
energy band being studied.  Second, if the energy dependence
of the modulations in the different bands accidentally resembles
that of the flaring, i.e., $\delta M_k \simeq d\ln F_k/d\ln
R_{\rm i,j} (\delta M_i - \delta M_j)$, then the modulations in
the corrected data set will have a greatly reduced amplitude,
possibly rendering them undetectable.

We have chosen a simple functional form to represent the
spectral-intensity changes in the different energy channels of the ASM
(also see Figure 3):
\begin{equation}
F_k(R_{\rm i,j}) = \alpha_k + \beta_k R_{\rm i,j}^{\gamma_k}  \,  .
\end{equation}
A table of parameters for the various ASM energy channels is
given for reference in Table 1.  For this particular functional form
we have
\begin{equation}
d\ln F_k/d\ln R_{\rm i,j} = \frac{(F_k-\alpha_k)\gamma_k}{F_k}  \, .
\end{equation}
Note that $d\ln F_k/d\ln R_{\rm i,j}$ is only approximately constant
through the flaring cycle and, in fact, varies by $\pm 7$\% to $\pm
20$\%, depending on the energy band.  Typical values of $d\ln F_k/d\ln
R_{\rm i,j}$ for each of the energy bands of SSC 2 are also listed in Table
1.

\section{Results}

We have applied the hardness-ratio-based correction procedure
described above (see, e.g., eq. 5) to each of the energy channels in
each of SSCs 1 and 2. The corrected data set for the C channel of SSC
2 is shown as an example in Figure 1c, while the hardness ratios used
in the procedure are shown in Fig. 1b.  A comparison of the corrected
data with the raw data indicates a dramatic reduction in source
variability (relative to mean source strength).  Corrected data sets
from the A and B channels of SSC 2, as well as from all three channels
from SSC 1, also have greatly reduced variability.

As was done for the raw data, the corrected data were binned into
0.01-d time bins and the mean value subtracted from those bins
containing data. The binned data sets were padded with zeros and then
transformed using the FFT.  Again, the resulting powers were
normalized to the mean power across the spectrum.  The results for the
C channel of SSC 2 are shown in Figure 2b.  Peaks near frequencies of
$\sim 1$ cycle d$^{-1}$, related to a window function frequency, are
now clearly evident in the PDS.

Amongst the other potentially interesting peaks in the frequency range
of 0.5 to 1.75 d$^{-1}$ is one at a frequency corresponding to a
period of $0.78893 \pm 0.00010$ d.  The normalized power reaches a
value of 11 close to this frequency, and for an assumed exponential
probability distribution, the probability of a statistical fluctuation
causing a peak this large in a particular frequency bin is $\sim 2
\times 10^{-5}$.  Given that there are roughly 2500 independent
frequencies in the range displayed in Fig. 2b, the probability of
finding a peak this large in any bin is $\sim 5\%$.  Although this is
not of great significance by itself, this frequency is actually of
{\em a priori} interest.  A blowup of this region of the PDS is shown
in Figure 4a.  The peak appearing in the SSC 2, channel C data is
clearly removed from the frequency corresponding to the GWL period,
but it does appear to be coincident with one of the other frequencies
identified in the analysis of GWL, i.e., the frequency corresponding
to a period of 0.78901 d.  In fact, the difference in frequency
between the GWL prime frequency and the frequency of the peak we see
corresponds to a period of $384 \pm 22$ days -- i.e., quite consistent
with a 1-year sideband of the GWL period at a frequency corresponding
to a period of 0.789014 d ($1.0/0.787313$ d $ -\ 1.0/365.25$ d $ =
1.0/0.789014$ d).

This peak at 0.78893 d appears with roughly equal amplitude in the A,
B, C and summed (A+B+C) bands (see Fig. 5).  Note that, while the data
in the three different bands are independent in terms of photon
counting statistics, the power in them is dominated by the variability
of Sco X-1 whose spectrum does not change by large factors.  Thus, the
different channels are all similarly affected by both the source
variability and the window function.  Moreover, the flaring in each of
the bands was removed utilizing the same hardness ratio $R_{\rm C,B}$
(not the same functions $F_k$, however).  Thus, the corrected data
sets used to produce the PDSs in Fig. 5 do not represent three
completely independent sets of observations.

The analysis procedure described above, with the appropriate hardness
ratio corrections, was also applied to the data from SSC 1.  The
results are shown in Figure 4b.  The normalized power in the bin that
we have identified as the likely orbital frequency of Sco X-1 is
5.7. This is not very statistically significant.  In fact, one expects
a peak this large by random chance somewhere within the frequency
range displayed in Figure 4.  Nonetheless, this peak at 0.78895 d is
the second highest among the $\sim 100$ independent frequencies in the
plot.  We take this as a {\em weak} confirmation that this particular
frequency is interesting.  The difference between the peak heights in
the transforms of data from the two different cameras is more or less
consistent with a signal that is of intermediate strength and
therefore at the margin of detectability in each.

We compared the results of the correction procedure described above
with other methods of reducing the power from the flaring behavior.
The other methods included (i) eliminating those measurements above a
specified intensity threshold chosen to be just above a baseline
value, e.g., 1050 cts s$^{-1}$ in the 1.5-12 keV band of SSC 2, and
(ii) eliminating those measurements with hardness ratios corresponding
to typical flaring values, i.e., $R_{\rm C,B} > 1.25$ (see Fig. 3).
In general, these other methods yielded smaller peaks close to a
period of 0.789014 d than did our correction procedure described above.
Combination of our correction procedure with alternative method (ii)
resulted in peaks which are comparable in strength to those found
without elimination of any measurements.

We carried out two further checks of our analysis of the data from SSC
2.  First, we transformed the window function corresponding to the
data set from this camera.  No interesting peaks were found near the
period of Sco X-1.  Second, we transformed different subsections of
the data to ascertain whether the amplitude of the orbital modulation
might be varying with time.  Unfortunately, with such a weak signal we
are able to say only that the modulation is consistent with having
been stronger during the first half of the {\it RXTE} mission.

We see no interesting harmonic peaks in the PDS at two or three
times the frequency of the peak that we have identified.

Next, we estimate the amplitude of the orbital modulation detected in
SSC 2.  Due to the fact that the data have been corrected by dividing
by a function of the hardness ratio, it is not obvious what the
Fourier amplitude or a fold, modulo the orbital period, will reveal
about the intrinsic amplitude or shape of the orbital modulation.
Therefore, we have chosen, somewhat arbitrarily, to calibrate the
amplitude scale in the {\em forward direction} by the following means.
We introduce a fictitious orbital modulation at a nearby frequency
$f_{cal} = 1.2600\ \rm{d}^{-1}$ with a $1\%$ amplitude in all three
energy channels by multiplying the actual data from SSC 2 by the
factor $1 + 0.01 \sin(2\pi f_{cal}t + \phi)$.  The results are shown
in the transforms displayed in Figure 5. From a direct comparison of
the ``calibration'' peak and the one at 0.78893 d (1.2675 d$^{-1}$),
we conclude that the orbital modulation amplitude lies in the range
$\sim 0.7\% - 1.4\%$. The higher end of this range allows for the
possibility that the modulations were present during only half the
time of the ASM observations.  This is quite consistent with the fact
that the peak at 0.78893 d appears broader than the calibration peak
by a factor of $\sim 2$.  Finally, we note that introducing equal
calibration amplitude signals into each energy band is equivalent to
assuming that the orbital modulations are colorless and that the third
term in curly brackets in equation (7) vanishes.

Under the assumption that the amplitudes of the orbital modulations
are not strongly dependent on the energy band, we have used a folded
light for the corrected SSC 2 data set to determine the time of
minimum X-ray intensity.  We find that these minima occur at times
$t$(X-ray minimum) $= \rm{MJD}\ 51334.93 + 0.789014E$ (MJD = Modified Julian
Date = Julian Date$\ -\ 2,400,000.5$).  The times of optical photometric
minimum are given by GWL as $t$(opt. minimum) $= \rm{MJD}\ 40080.63 +
0.787313E$.  On 1999 June 6 (MJD 51335), the GWL ephemeris predicts
$t$(opt. minimum) $= \rm{MJD}\ 51335.26$ which is 0.33~d or 0.42
cycles after X-ray minimum.  Thus, X-ray minimum is coincident with
optical photometric maximum within the uncertainty in the phases of
the cycles ($\delta \phi \sim 0.1$).  However, we point out that
equation (7) allows for the possibility of a phase reversal (i.e., by
$180^\circ$) in the X-ray minimum of a light curve formed from the
{\em corrected} data, depending on the details of the hardness ratio
behavior of the orbital modulations.

\section{Discussion}

We find evidence for a periodicity in the X-ray flux from Sco X-1 with
a period of $0.78893 \pm 0.00010$ d, and identify this as likely
caused by some physical effect producing modulations of the flux at
the orbital period.  We do not find good evidence for variability at
the 0.787313 d period.  In fact, GWL found a periodicity at 0.78901 d
that is consistent with the period we find in the ASM data, but since
they found that periodicity to be $\sim 20$\% weaker than that at
0.787313 d, they concluded that the latter was the orbital period while
the former was the 1-year sideband produced by the observational
window function.  We suggest rather that the true orbital period is
0.78901 d and that the 0.787313 d period is the 1-year sideband.
Unfortunately, we are unable to evaluate quantitatively the likelihood
of the sideband appearing stronger in the photometric data than the
true modulation frequency since we do not have access to the plate
stack data used by GWL.

We note, however, that ASM observations of Sco X-1 are generally
available for more than 10 months each year while most ground-based
optical observations of Sco X-1 were likely taken within a couple of
months of early June when Sco X-1 is nearly opposite from the Sun.
Thus, it is likely that the window function of the optical observations
analyzed by GWL would have strong components at a 1-year period.  We
further note that a modulation with a period of 0.78901 d that is in
phase with a modulation at 0.787313 d as described by GWL in early
June would be difficult for optical observers to distinguish without
observations at other times of the year.  For example, the two
oscillations would be out of phase by only 0.08 cycles in early May or
early July.  At even earlier or later dates, the phases of the two
modulations would gradually drift apart until they came back together
(with a difference of 1 cycle) after one year.  We can therefore
speculate that there were a limited number of optical observations in
the set used by GWL that could have differentiated between the two
periods.

The difference between the two periods could well have been missed by
independent observers who confirmed the GWL period. Although the
optical spectroscopic results of Cowley \& Crampton (1975) do support
the presence of a periodicity, they did not have the frequency
resolution to distinguish between the GWL period and a one year
sideband.  We also expect that the independent measurements of LaSala
\& Thorstensen (1985) cannot be used to distinguish the GWL period
from one which differs by 1 cycle yr$^{-1}$.

Due to the small observed amplitude of the X-ray orbital modulation in
Sco X-1 we are unable to investigate either the detailed shape of the
light curve or the spectral character of the modulations.  However, it
is clear from the lack of higher harmonics in the FFTs and the
similarity in amplitudes of the peaks in Fig. 5 that the light curve
has no prominent sharp structure nor is it highly energy
dependent. Specifically, the modulation cannot be due to photoelectric
absorption in unionized matter which would have a much larger effect
in the 1.5-3 keV channel than in the 5-12 keV band.  As noted in the
previous section, the phase of X-ray minimum appears to be 0.42 cycles
earlier than that in the optical.  Based on any simple picture for
scattering or absorption in the system, one would expect the optical
and X-ray minima to be roughly coincident, unless the outgoing
radiation is affected by structure on the rim of the accretion disk
which is fixed asymmetrically in the rotating frame of the binary.  At
the low inclination angle that we are likely to be viewing Sco X-1
(e.g., $i \simeq 30^\circ$, Steeghs \& Casares 2002) we would not
expect this type of effect.  Thus, if our orbital period is correct,
we have no ready explanation for the phase of X-ray minimum.

In general, the orbital modulation could be due to attenuation of the
beam, i.e., scattering or absorption along the line of sight, or to
scattering into the line of sight.  Possible sources of the scattering
and/or absorption could include a stellar wind of the companion, the
surface of the companion, a wind from the accretion disk, or an
accretion disk corona (see, e.g., Priedhorsky \& Holt 1987). If the
modulation takes place near the accretion disk, then some asymmetry
around the azimuth of the disk, fixed in the rotating frame of the
binary, must be present.

If the donor star serves as a source of scattered radiation (see
Priedhorsky \& Holt 1987), then we can estimate the maximum effect by
computing the solid angle subtended by the donor star.  The fractional
solid angle is given by $\sim 0.053 (1+q)^{-2/3}$, where $q$ is the
ratio of the mass of the neutron star to that of the donor star.  For
plausible masses (see Steeghs \& Casares 2002) this fraction is $\sim
2$\%.  In order for a $\sim 1$\% orbital modulation to be due to
X-rays scattered into the line of sight from the companion star, this
would require an implausibly high X-ray albedo of $\sim 0.5$.

The accretion disk might afford a larger solid angle for intercepting
X-rays and scattering them into the line of sight.  For example, an
accretion disk with a half opening angle of $10^\circ$ would subtend
at the neutron star a fractional solid angle of $\sim$17\%.  For
reasonable X-ray albedos of $\sim 10$\% for glancing incidence (see
below), this could result in some 1.7\% of the X-rays emanating from
the Sco X-1 system being scattered.  If there is some asymmetry around
the disk that yields about a factor of two more scattering at some
azimuths than others, this could produce a net orbital modulation of
$\sim 1$\%.

The X-ray albedo is $\sim 0.5 \sigma_e/\sigma_{pe}$, where $\sigma_e$
is the angle-averaged electron scattering cross section, and
$\sigma_{pe}$ is the energy-dependent photoelectric absorption cross
section appropriate for the composition and ionization state of the
scattering medium.  The exact coefficient depends on the geometry
experienced by the incident and emerging X-rays.  For material that is
of solar composition and nearly unionized, the albedo should scale
roughly as $E^{8/3}$; it should be a strongly increasing function of
photon energy, $E$, that reaches $\sim 6$\% in the middle of the ASM C
band.  For more highly ionized plasmas, the albedo can be
substantially higher and less energy dependent. The latter property
would be relevant in trying to model the nearly energy-independent
amplitude we find for the orbital modulation.

A small number of X-ray sources that exhibit orbital modulation have
been classified as ``accretion disk corona'' (ADC) sources, e.g., 4U
1822--37 and 4U 2129+47 (see, e.g., White \& Holt 1982).  The ADC
sources are supposedly viewed nearly edge on, with the direct path of
X-rays from the neutron star blocked by an accretion disk.  In this
case, most of the X-rays detected at the Earth may be scattered by a
hot corona located above and below the inner parts of the accretion
disk.  The orbital modulation of the observed X-ray flux may then be
produced by a thickened disk rim that occults the corona to differing
degrees as a function of orbital phase \citep{WH82}.  The modulation
is typically large enough in amplitude, e.g., $\sim$20\% of the
average flux, to be easily detectable even outside of the relatively
narrow partial eclipse.  However, we cannot say whether similar
systems observed at much lower inclinations would show modulations at
the $\sim$ 1\% level, because (1) the neutron star would be visible
and (2) the disk rim would not occult the corona.

Finally, to help establish the correct orbital period in Sco X-1 we
urge optical astronomers to search for modulations in the light from
or radial velocity of this system as far from June as practicable.

Acknowledgements

The authors thank Bill Liller, Ron Remillard, and Ned Wright for very
helpful discussions, and Arnout van Genderen and Jan Lub for providing some
of their archival optical data from Sco X-1. Partial support for
this research was provided by NASA Contract NAS5-30612 and NASA Grant
NAG5-9189.

\begin{deluxetable}{lcccc}
\tablewidth{0pt}
\tablecaption{Intensity vs. Hardness Ratio Parameters for SSC 2\label{table1}}
\tablehead{
\colhead{Energy Band} &
\colhead{$\alpha_k$} &
\colhead{$\beta_k$} &
\colhead{$\gamma_k$} &
\colhead{$d\ln F_k/d\ln R_{\rm C,B}$\tablenotemark{a}}
}
\startdata
1.5-3 keV (A) & 140 & 100 & 2.0 & 1.0 \\
3-5 keV (B) & 113 & 141 & 2.0 & 1.3 \\
5-12 keV (C) & 50 & 210 & 2.5 & 2.2 \\
1.5-12 keV (sum) & 500 & 280 & 3.0 & 1.5 \\
\enddata
\tablenotetext{a}{$R_{\rm C,B}$ is the 5-12 keV band to 3-5 keV band
hardness ratio, and $d\ln F_k/d\ln R_{\rm C,B}$ is given for $R_{\rm
C,B} = 1.2$.}
\end{deluxetable}

\begin{figure}
\figurenum{1}
\epsscale{0.9}
\plotone{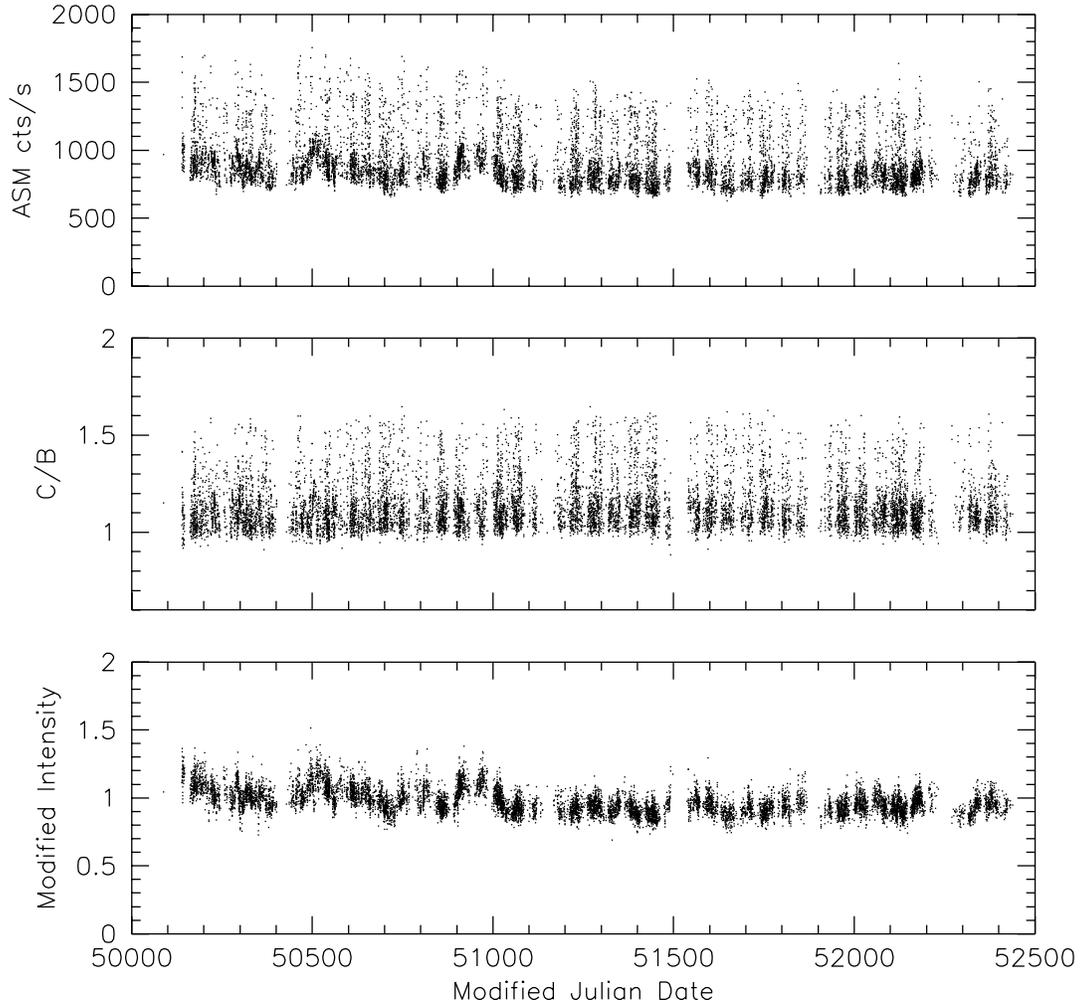}
\caption{
(top) The 1.5-12 keV intensity of Sco X-1 from observations
made with Scanning Shadow Camera 2 of the {\it RXTE} ASM.  The
intensity is the average over each 90 s observation and is given as
the count rate that would have been recorded with the source at the
center of the field of view of SSC 1 in early 1996.  In these units,
the intensity of the Crab Nebula is 75 counts s$^{-1}$.  (middle) The
hardness ratio, defined as $I(5-12~{\rm keV})/I(3-5~{\rm keV})$, derived
for each SSC 2 observation of Sco X-1. (bottom) Corrected intensity of
Sco X-1 from SSC 2 observations (see text).
\label{fig1}}
\end{figure}

\begin{figure}
\figurenum{2}
\epsscale{0.9}
\plotone{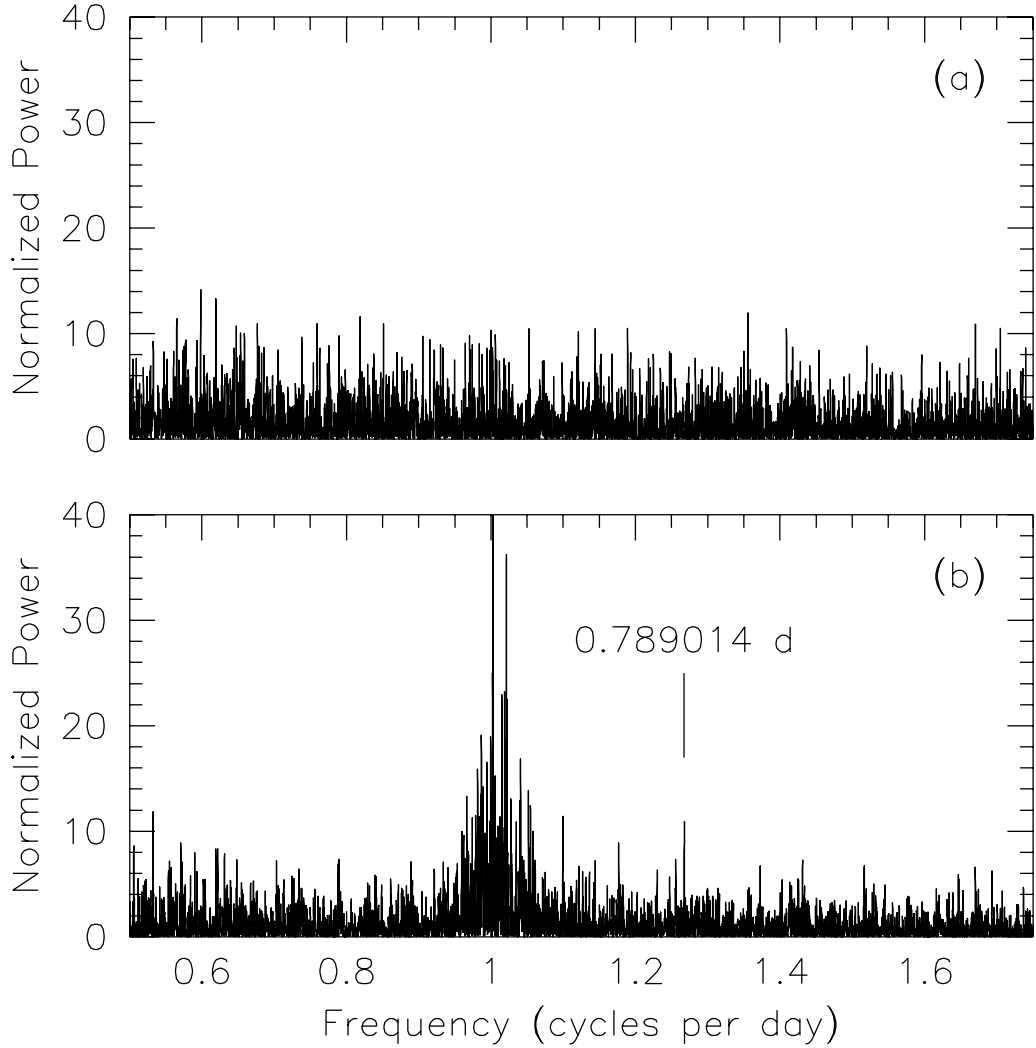}
\caption{
(a) Normalized power density spectrum of the uncorrected SSC
2 C band (5-12 keV) data. (b) Normalized power density spectrum of the
corrected SSC 2 C band data.  A mark indicates the frequency
corresponding to a period of 0.789014 days near which there is a peak
of interest.
\label{fig2}}
\end{figure}

\begin{figure}
\figurenum{3}
\epsscale{0.9}
\plotone{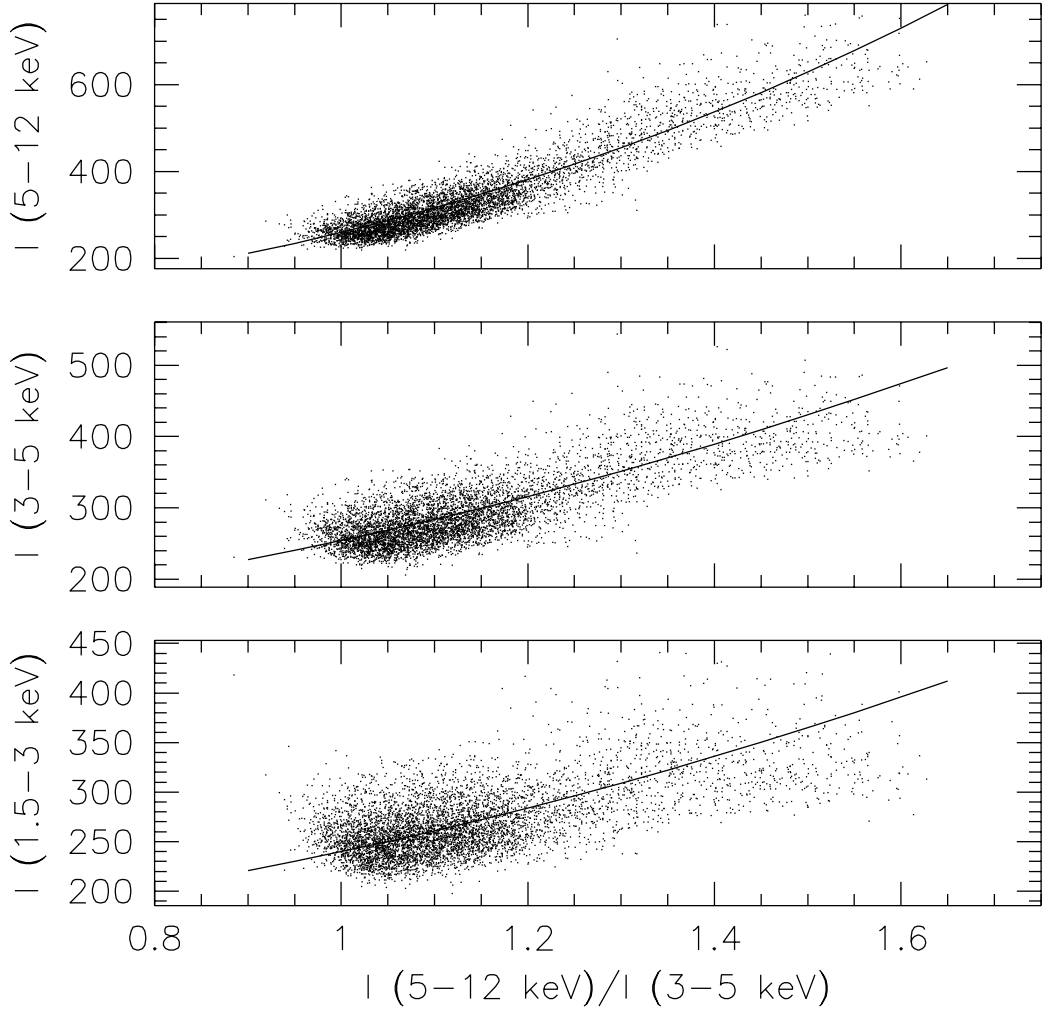}
\caption{
(top) SSC 2 C band (5-12 keV) intensities plotted vs. the C
band to B band hardness ratio, $R_{\rm C,B}$.  (middle) SSC 2 B band
(3-5 keV) intensities vs. the hardness ratio.  (bottom) SSC 2 A band
(1.5-3 keV) intensities vs. the hardness ratio.  In all three panels,
the solid curve is a model of the mean relation between the intensity
and hardness ratio (see eq. 8 and Table 1).
\label{fig3}}
\end{figure}

\begin{figure}
\figurenum{4}
\epsscale{0.9}
\plotone{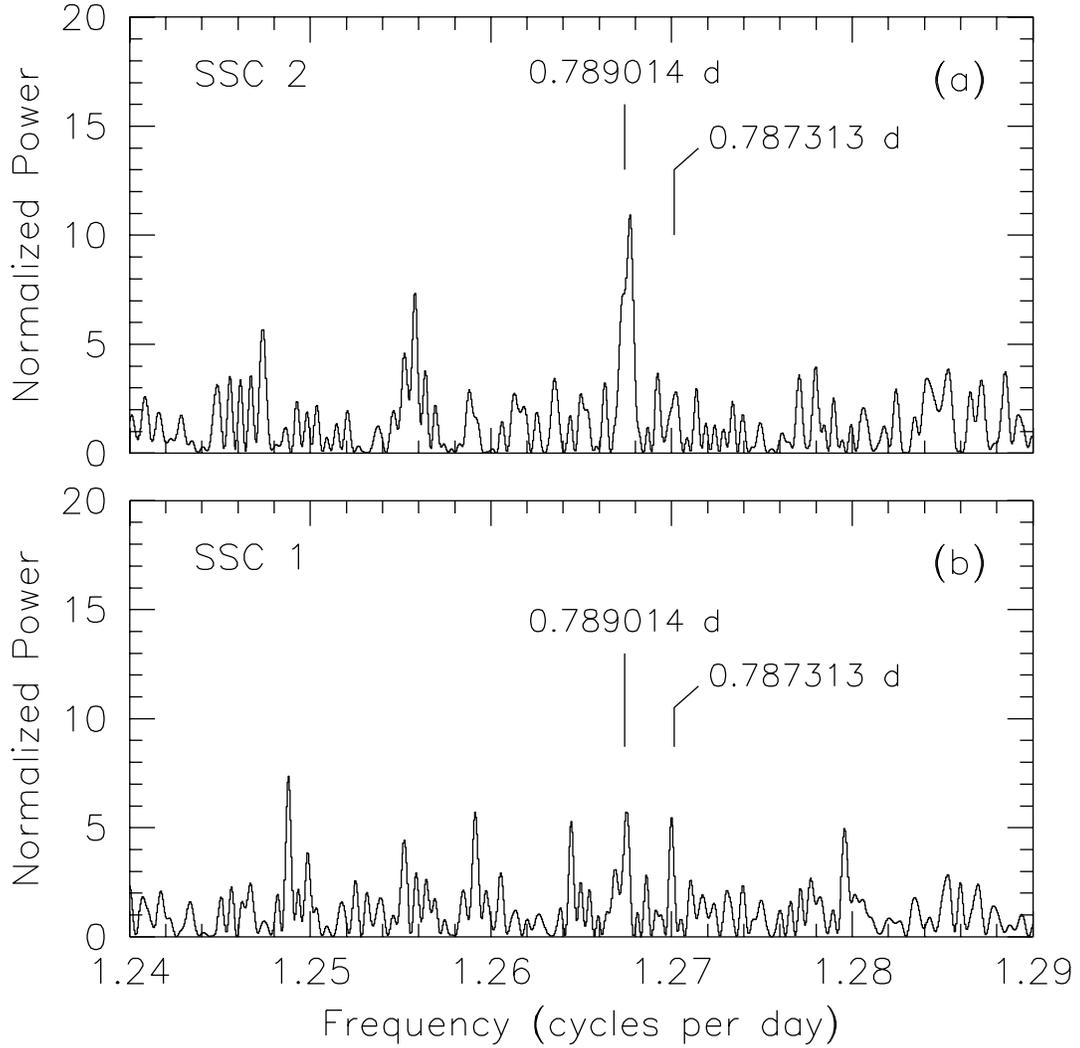}
\caption{
(a) Normalized power density spectrum of the corrected SSC 2
C band data in the frequency region around the orbital period of Sco
X-1.  Marks indicate frequencies corresponding to periods of 0.789014
days and 0.787313 days (GWL's orbital period).  (b) Normalized power
density spectrum of the corrected SSC 1 C band data.
\label{fig4}}
\end{figure}

\begin{figure}
\figurenum{5}
\epsscale{0.9}
\plotone{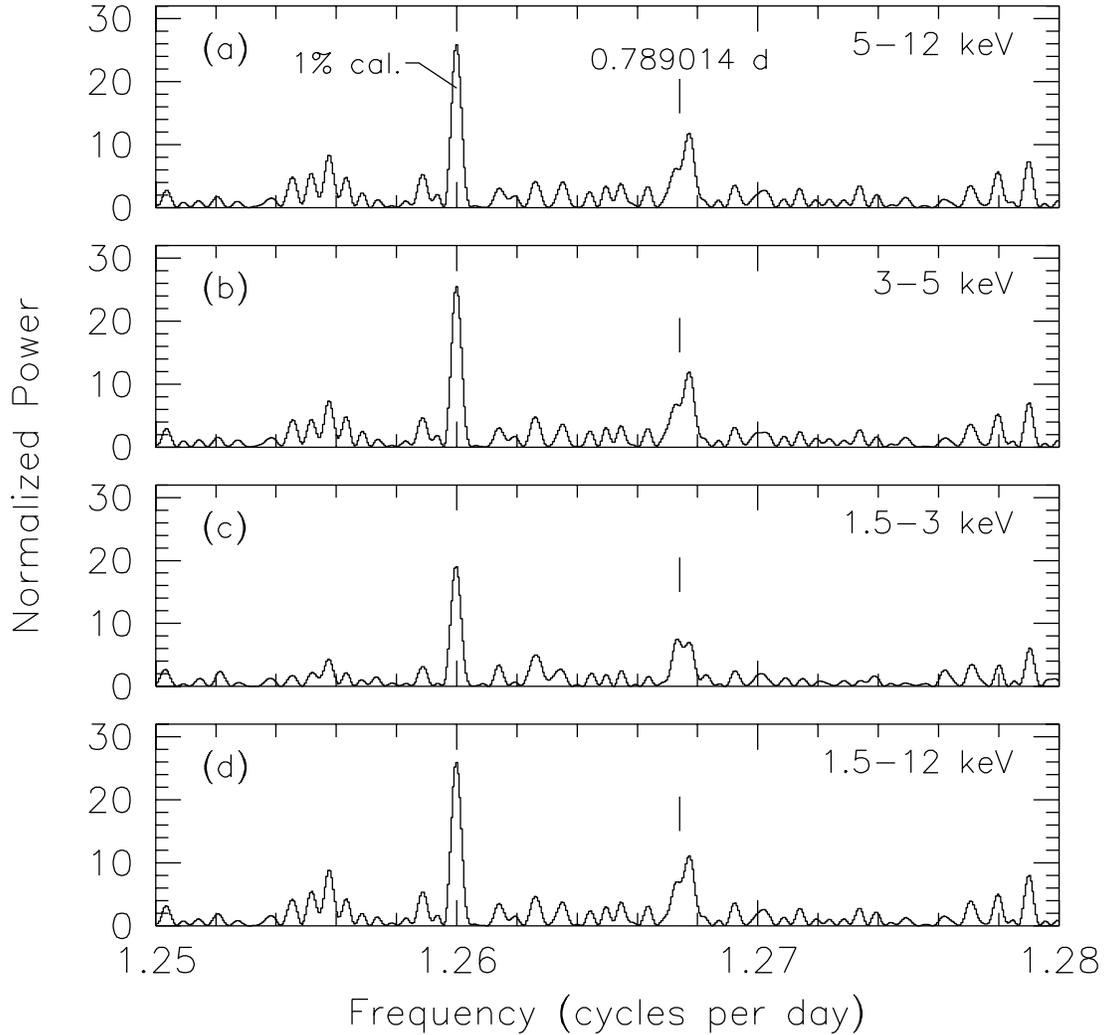}
\caption{
(a) Normalized power density spectrum of the corrected SSC 2
C band data to which a sinusoidal signal at a frequency of 1.26 cycles
d$^{-1}$ with a semi-amplitude of 1\% (rms amplitude 0.7\%) was added
in software.  A mark indicates the frequency corresponding to a period
of 0.789014 days. (b, c, d) Normalized power density spectra for the
B, A, and sum bands with 1\% calibration signals added.
\label{fig5}}
\end{figure}

\end{document}